\documentclass{aastex631}

\usepackage{tablefootnote}

\shorttitle{ICL analysis of fossil groups}
\shortauthors{de Oliveira et al.}

\begin{document}

\title{The intracluster light analysis of the most evolved systems of galaxies: fossil groups}

\email{nicolasoliveira@on.br}

\author[0000-0002-8742-0643]{N\'icolas O. L. de Oliveira}
\affiliation{Observat\'orio Nacional, Rua General Jos\'e Cristino, 77 - Bairro Imperial de S\~ao Crist\'ov\~ao, Rio de Janeiro, 20921-400, Brazil}

\author[0000-0002-6090-2853]{Yolanda Jim\'enez-Teja}
\affiliation{Instituto de Astrof\'isica de Andaluc\'ia--CSIC, Glorieta de la Astronom\'ia s/n, E--18008 Granada, Spain}
\affiliation{Observat\'orio Nacional, Rua General Jos\'e Cristino, 77 - Bairro Imperial de S\~ao Crist\'ov\~ao, Rio de Janeiro, 20921-400, Brazil}

\author[0000-0003-1477-3453]{Renato A. Dupke}
\affiliation{Observat\'orio Nacional, Rua General Jos\'e Cristino, 77 - Bairro Imperial de S\~ao Crist\'ov\~ao, Rio de Janeiro, 20921-400, Brazil}
\affiliation{Department of Astronomy, University of Michigan, 311 West Hall, 1085 South University Ave., Ann Arbor, MI 48109-1107}
\affiliation{Eureka Scientific, 2452 Delmer St. Suite 100,Oakland, CA 94602, USA}

\author[0000-0002-7272-9234]{Eleazar R. Carrasco}
\affiliation{International Gemini Observatory/NFS National Optical Infrared Research Laboratory, Casilla 603, La Serena, Chile}

\author[0000-0002-6610-2048]{Anton M. Koekemoer}
\affiliation{Space Telescope Science Institute, 3700 San Martin Drive, 
Baltimore, MD 21218, USA}

\author[0000-0002-3886-1258]{Yuanyuan Su}
\affiliation{Department of Physics and Astronomy, University of Kentucky, 505 Rose Street, Lexington, KY 40506, USA}

\author[0000-0001-7299-8373]{Jose Manuel Vilchez}
\affiliation{Instituto de Astrof\'isica de Andaluc\'ia -- CSIC, Glorieta de la Astronom\'ia s/n, E--18008 Granada, Spain}

\author[0000-0003-4307-8521]{Jimmy A. Irwin}
\affiliation{Department of Physics and Astronomy, University of Alabama, Box 870324, Tuscaloosa, Alabama, 35487, USA
}

\author[0000-0002-3031-2326]{Eric D. Miller}
\affiliation{Kavli Institute for Astrophysics and Space Research, Massachusetts Institute of Technology, 70 Vassar St, Cambridge, MA 02139, USA}

\author{Lucas E. Johnson}
\affiliation{University of West Alabama, 100 US-11, Livingston, AL 35470, USA}

\begin{abstract}
We present the analysis of the intracluster light (ICL) in three fossil groups (FG), RX
J085640.72+055347.36, RX J1136+0713, and RX
J1410+4145, at z $\sim$ 0.1. We used two optical broad-band filters, F435W and F606W, observed with the Hubble Space Telescope and spectroscopic data obtained with the Gemini Multi-Object Spectrograph to generate the ICL maps and measure the ICL fraction using CICLE, an algorithm developed to disentangle the ICL from the light of galaxies. We found ICL fractions of $9.9\%-14.4\%$, $3.8\%-6.1\%$, and $4.7\%-10.7\%$ for RXJ0856, RXJ1136, and RXJ1410, respectively. 
This behavior is not consistent with the presence of the ICL fraction excess previously observed in merging clusters and also inconsistent with the constant ICL fraction distribution characteristic
of relaxed systems, although the values found are within the typical range expected for the latter. Instead, they show a significantly increasing trend with wavelengths over $\sim3800 - 5500$ \AA, indicating that fossil groups are indeed old and undisturbed systems, even compared with regular relaxed clusters. 
\end{abstract}

%

\section{Introduction} \label{sect:introduction}

Fossil groups (FG) are classically defined as those systems which satisfy two conditions: 1) they are dominated by a single elliptical galaxy which is at least 2 magnitudes brighter than the second ranked galaxy member located within half the virial radius $R_{200}$, and 2) they have an extended halo of X-ray emitting gas with luminosity $L_{\text{X,bol}}\geq 10^{42} h_{50}^{-2}$ erg s$^{-1}$ \citep{ponman1994,jones2003}. The traditional scenario of formation of FGs describes that the brightest group galaxy (BGG) must have accreted relatively early its massive satellite galaxies, which approached the BGG through dynamical friction likely following orbits with low angular momentum \citep[e.g.,][]{vonbenda2008}. Given the long times involved in dynamical friction and the absence of X-ray substructures observed in FGs, this scenario implies that FGs are old systems which stayed unperturbed for a long time and should be at the end of their merging tree \citep{vikhlinin1999,jones2000}. Indeed, BGGs in FGs are morphologically different than those in non-FGs, suggestive of different formation histories \citep{chu2023,habib2006}. Moreover, FGs usually present high concentration parameter values and high X-ray flux 
concentration \citep{khosroshahi2007, santos2008}. As the concentration parameter correlates with the formation epoch of the groups and clusters of galaxies under the currently accepted $\Lambda$CDM standard cosmological model \citep{wechsler2002}, this would also imply an early formation time for FGs. The high X-ray flux 
concentration would imply well-developed cool cores  \citep[e.g.,][]{hudson2010}.\\

However, the lack of cool cores observed in many of these systems challenges the standard explanations for their evolution. Furthermore, BGGs in FGs and non-FGs share similar stellar populations, which is difficult to explain if they followed different evolutionary paths \citep{chu2023}. This raises the question of the purity of the FG samples or, equivalently, whether all systems classified as FGs are truly highly evolved systems.
On the one hand, the magnitude gap criterion alone is prone to draw misleading conclusions since it is affected by both physical and observational systematics. Some photometrically selected FGs have been shown to be regular systems when spectroscopic measurements have been available \citep{aguerri2011}. On the other hand, it has been shown that some systems can satisfy (at least temporarily) the FG criteria without having the properties of highly evolved systems. \\

Recently, \cite{dupke2022} applied the analysis of the intracluster light (ICL hereafter)  to a classic FG at $z \sim 0.112$, RX J100742.53+380046.6 (RXJ1007 hereafter), to assess its dynamical state and to estimate the epoch of its last merger. The ICL is defined as the diffuse, low-surface-brightness component of clusters consisting of stars detached from their host galaxies. We here use the term ICL to denote also the analogous component for galaxy groups.  
The ICL is essentially composed by the stellar remnants of the galactic interactions during the assembly of the cluster, so it encodes information about the accretion history of the system and its dynamical state. Indeed, \cite{jimenez-teja2018} showed that measurements of ICL fractions, defined as the flux ratio of the ICL to the total cluster light, can trace the dynamical state of clusters and identify merging against relaxed systems. \cite{dupke2022} measured the ICL fractions of RXJ1007, finding that the system was highly relaxed, more so than any other cluster where the same analysis had been conducted. 
In this work, we 
apply a similar ICL analysis to three additional FGs to check 
whether they are indeed extremely evolved systems and, if so, to identify a possible pattern in their ICL fractions. Since all our 3 FG candidates have $z \sim 0.1$, they are directly 
comparable to RXJ1007. \\

This paper is organized as follows: Sect. \ref{sect:data} describes the optical imaging and the spectra used in the analysis of the ICL for all FGs. Sect. \ref{sect:analysis} describes the processing steps to generate the ICL maps and measure the ICL fraction. In Sect. \ref{sect:discussion}, we discuss our results of the ICL and, finally, we summarize and draw our conclusions in Sect. \ref{sect:conclusions}. Throughout this work, we assume a standard $\Lambda$CDM model with $H_0=70$ km s$^{-1}$ Mpc$^{-1}$, $\Omega_m=0.3$, and $\Omega_{\Lambda}=0.7$. All magnitudes are expressed in the AB system. \\


\section{Data} \label{sect:data}

In this work, we analyze three FGs: RX J085640.72+055347.36, RX J1136+0713, and RX J1410+4145, (RXJ0856, RXJ1136, and RXJ1410 hereafter, respectively). Their coordinates and redshift are listed in Table \ref{table:clusters}.
These systems were previously identified as FG by \cite{miller2012}, who found 12 systems that satisfied the empirical definition of FGs in optical and X-ray observations. The authors used the maxBCG cluster catalog \citep{koester2007} of over 17,000 optically selected red-sequence clusters in the redshift range $0.1 < z < 0.3$, based on precise photometric redshifts from the Sloan Digital Sky Survey DR4 \citep[SDSS,][]{york2000,adelmanmcCarthy2006}. From that sample, we obtained for four of them optical imaging with the Hubble Space Telescope (HST) and spectra with Gemini. One of them, RXJ1007, has been analyzed by \cite{dupke2022} and we show the ICL analysis of the other three here.\\

\subsection{HST imaging} \label{sect:data:HST}

\begin{figure}
\includegraphics[width=1.\textwidth]{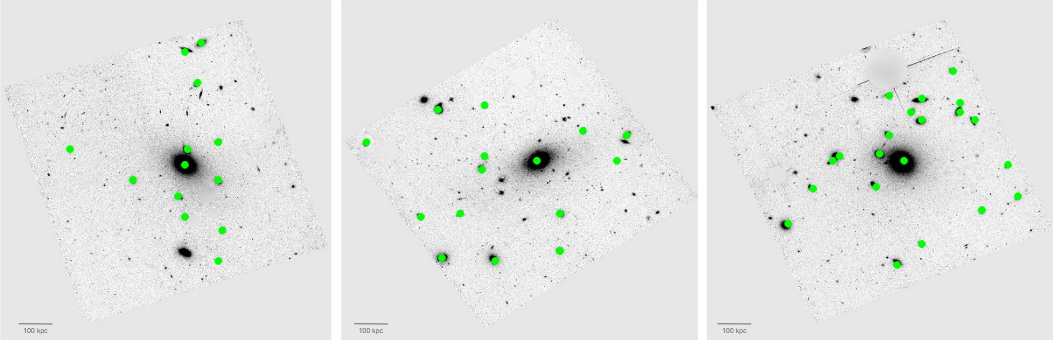}
\caption{Original images of RXJ0856 (left), RXJ1136 (center), and RXJ1410 (right) in the F606W filter and their corresponding galaxy members in green circles. Saturated stars are masked in white circles. North is up, east is left.} \label{fig:fgs}
\end{figure}

The three FGs (Figure \ref{fig:fgs}) were observed with the HST Advanced Camera for Surveys (ACS/WFC) in the broad-band filters F435W and F606W within the HST Program \#15671 (PI: Dupke). Each cluster was observed for a total of 3 orbits, consisting of 1 orbit with the F435W filter and 2 orbits with the F606W filter. Each orbit was split into 4 exposures obtained using the pre-defined ACS-WFC-DITHER-BOX pattern with default parameters optimized to provide a combination of integer and half-pixel sampling. Thus, for each FG target, the single orbit of F435W data consists of 4 dithered exposures, each with an exposure time of 505 seconds, and the 2 orbits of F606W data consist of 8 dithered exposures, each with an exposure time of 509 seconds.

All the exposures were calibrated using the STScI ACS calibration software to account for instrumental effects including bias, dark current, flat fielding, bad pixels, CTE correction, and hot pixels, and were processed following the approaches first described in \cite{koekemoer2003,koekemoer2011} to account for geometric distortion, perform astrometric alignment and cosmic ray rejection, and produce final combined mosaics at pixel scales of 0.03\arcsec and 0.06\arcsec, covering an area approximately 200 arcsec across, and all aligned to a common pixel grid.

\subsection{Gemini spectra} \label{sect:data:Gemini}

We obtained spectroscopic data of the three FGs with the Gemini Multi-Object Spectrograph \citep[hereinafter GMOS, ][]{hook2004} mounted at the Gemini North telescope in Hawaii, in queue mode. The GMOS multi-object observations (MOS) for RXJ0856 were carried out between 2021 January 13 and 18 UT (Program ID: GN-2020B-Q-302, PI: Batalha) and for RXJ1136 and RXJ1410, between 2024 February 06 and 2024 March 16 UT (Program ID: GN-2024A-Q-107, PI: Dupke), during dark time, under clear skies and with a seeing between 0.7\arcsec and 1\arcsec. Two masks per FG were designed. Spectra were acquired using the B600 (RXJ0856) and the B480 gratings (RXJ1136 and RXJ1410) centered at 5500 \AA, using 1\arcsec~slitest and $2 \times 2$ binning.  
Offsets of 100$^{\circ}$ \AA~ toward the blue and the red were applied between exposures to cover the gaps between the CCDs. At each wavelength setting, spectroscopic flats and CuAr comparison lamp spectra were taken before or after each science exposure. All spectra were flux-calibrated using the spectrophotometric standard stars BD+52 913 (RXJ0856) and G191B2B (RXJ1136 and RXJ1410)
observed with the same instrument setup as the science images, but on different nights
and under different observing conditions. Therefore, only a relative flux calibration of the science spectra is provided.\\

The science spectra and corresponding calibrations were reduced using the Gemini GMOS package version 1.16. In summary, the science and calibration exposures were overscanned, bias-subtracted, and trimmed. The two-dimensional science exposures were then flat-fielded, wavelength calibrated, distortion corrected, and extracted to a one-dimensional format. The final wavelength solution has an average rms of $\sim 0.2$ \AA. The resolution of the extracted one-dimensional spectra is $\sim 5.8$ \AA~ (measured from the sky lines FWHM), with a dispersion of $\sim 1$ \AA~ (B600 grating) $\sim 1.25$ \AA~ pixel$^{-1}$ (B480 grating), covering a wavelength interval between $\sim 4000$ and 8000 \AA. \\

We determined the redshifts of the galaxies observed with GMOS using the programs implemented in the RV package inside IRAF. All spectra were cross-correlated with four high signal-to-noise (S/N) templates using the program FXCOR. The redshift errors were estimated based on the R statistic value of \cite{tonry1979}. For galaxies with obvious
emission lines, a line-by-line Gaussian fit was employed using the routine RVIDLINES. The errors of the measurements were estimated using the residual of the average redshift shifts of all measurements provided by the program. We were able to determine the redshifts for 170 galaxies included in the GMOS masks. Of these, 52 correspond to RXJ0856, 54 to RXJ1136, and 64 to RXJ1410. \\

The physical area covered for each of the FG with the GMOS MOS observations is summarized in Table \ref{table:clusters}. We increased the number of galaxies by adding galaxies with spectroscopic redshifts that were not included in our MOS observations using the SDSS DR18 database \citep{almeida2023}. The search was limited to galaxies with redshifts between $0.08 < z < 0.12$ and within half virial radius. The GMOS sample is then supplemented with 38 additional galaxies compiled from SDSS DR15 and DR18 (19 galaxies for RXJ0856, 9 for RXJ1136, and 10 for RXJ1410). Thirteen of these galaxies 
had redshift estimations in common with GMOS. We checked that the redshifts obtained with GMOS agreed well with those redshifts in the SDSS database, with a mean difference of $-3 \pm 4$ km s$^{-1}$.

\begin{table*}
\centering\caption{Coordinates of the three candidates for FGs (plus RXJ1007), details of the GMOS observations (observed time and physical area covered), and dynamical parameters derived from them.}
\begin{tabular}{lccccccccc}
\hline\hline
Cluster & R.A. & Dec & Physical area & $N_{\text{gal}}$ & $N_{\text{mem}}$\tablenotemark{a} & $\overline{z}$ & $\sigma_{los}$ & $R_{200}$ & $M_{200}$ \\ 
& & & (Mpc$^2$) & & & & (km s$^{-1}$) &  (Mpc) & ($10^{14} \, M_{\odot}$)\\
\hline
RXJ0856 & 08:56:40.72 & 05:53:47.36 & $0.85 \times 0.64$ & 52 & 28+9 & $0.093705\pm0.000357$ & $587 \pm  85$ & $1.40 \pm 0.20$ & $1.42 \pm 0.57$\\
RXJ1136 & 11:36:23.72 & 07:13:37.52 & $1.07 \times 0.76$ & 54 & 33+4 & $0.102788\pm0.000339$& $538 \pm  80$ & $1.28 \pm 0.19$ & $1.11 \pm 0.11$ \\
RXJ1410 & 14:10:04.19 & 41:45:20.88 & $0.98 \times 0.70$ & 64 & 37+5 & $0.093531\pm0.000360$ & $602 \pm  124$ & $1.44 \pm 0.17$ & $1.52 \pm 0.27$ \\
RXJ1007 & 10:07:42.53 & 38:00:47.50 & $2.95 \times 2.95$ & 98 & 20+26 & $0.111834\pm0.000480$ & $570\pm56
$ & $1.35\pm0.13$ & $1.30\pm0.35$ \\
\hline
\end{tabular}
\tablenotetext{a}{Number of members identified from the GMOS data plus those with SDSS spectroscopic information.}
\label{table:clusters}
\end{table*}


\section{Analysis} \label{sect:analysis}

We generate the ICL maps of the three candidates to FGs using the CHEFs Intracluster Light Estimator \citep[CICLE, ][]{jimenez-teja2016}, an algorithm especially designed to remove the galactic light and isolate the ICL. CICLE fits the galaxies and removes the models to leave the ICL alone. The models are created using orthonormal bases called CHEFs \citep{jimenez-teja2012}, built from Chebyshev rational functions and Fourier series. The only exception is the brightest cluster galaxy, whose modeling is particularly complex because its spatial distribution coincides in projection with that of the ICL. In order to delimit the BGG, we compute a curvature map of the BGG+ICL projected surface, which shows how the slope of this composite surface changes in each pixel. Those pixels where the curvature changes most indicate the transition from the BGG to the ICL. Finally, the ICL in the BGG-dominated region is interpolated from the outskirts following the slope and shape of the ICL-dominated region. It is important to note that we have discarded the pixels from saturated stars and from the borders of all images to avoid any contamination and/or bias due to artificial light. We refer the reader to \cite{jimenez-teja2016}, \cite{jimenez-teja2018} and \cite{jimenez-teja2021} for further details about CICLE and ICL contamination sources. A recent study compared CICLE and other techniques of ICL analysis by testing them against simulations \citep{brough2024}, concluding that CICLE's results were the most consistent with simulations and with the smallest scatter. We found that the BGG-ICL transition radius for the three FG candidates is in the range $\sim60-80$ kpc, making them comparable with the results found for their mock images of galaxy clusters at $z \sim 0$. Our results also agree with the findings of \cite{chu2023}, who reported no significant difference in the sizes of BGGs in FGs 
compared to those in non-fossil systems. \\

Our aim is to calculate the ICL fraction with the HST data, which is defined as the ratio between the ICL flux and that of the total cluster, i.e., ICL plus galaxies. Thus, we need to previously identify the cluster members with the available spectroscopic data. The number of member galaxies of all FGs are estimated using the robust bi-weight estimators of central location ($C_{BI}$) and scale ($S_{BI}$) of \cite{bee90}, using 
an iterative procedure and applying a 3-$\sigma$ clipping algorithm to remove outliers. The majority of member galaxies are located in the region defined by the  colour-magnitude diagram (CMD) relation for early-type galaxies in clusters; the Red Cluster Sequence. 
The best estimates of the location ($\overline{z}$), scale ($\sigma_{los}$), number of members ($N_{mem}$), virial radius ($R_{200}$), and mass ($M_{200}$) of the three clusters are shown in Table \ref{table:clusters}. The dynamical mass of the FGs was calculated using the $\sigma-M_{200}$ scaling relation of \cite{munari2013} obtained from zoomed-in hydrodynamical simulations of dark matter halos calibrated using dark matter particles and taking into account prescriptions for cooling, star formation, and active galactic nuclei feedback. The errors in $M_{200}$, $R_{200}$, and the redshifts were estimated using the bootstrap technique with 10,000 realizations. All errors quoted in Table \ref{table:clusters} are at the 68\% confidence level. The final number of member galaxies is 37 for both RXJ0856 and RXJ1136, and 42 for RXJ1410. \\

Finally, we calculate the total ICL fractions, i.e., using the whole region where the ICL is detected. The final values and the radius where these fractions are measured are listed in Table \ref{table:ICLf}. To assess whether the observed increase in the ICL fraction measured from F435W to F606W could be artificially driven by the difference in surface brightness limits between the filters, we applied a surface brightness cut to the F606W data, matching the limiting depth of the F435W images. After recalculating the ICL fractions using this artificially limited dataset, we found that the resulting values differ by less than $1\%$ from the original F606W measurements, well within the error bars in most cases. This small change confirms that the higher ICL fraction observed in F606W is not a result of a different surface brightness depth, 
thus supporting a physical origin for the different ICL fractions found at different wavelengths in FGs. 

\begin{table}
\centering \caption{ICL fraction, equivalent radius of the ICL, and limiting depth of the images for the three candidate for FGs (plus RXJ1007), as yielded by  CICLE.}\label{table:ICLf}
\begin{tabular}{lcccccc}
\hline\hline
Cluster & \multicolumn{3}{c}{F435W} & \multicolumn{3}{c}{F606W} \\
\hline
& ICL fraction & r &  Surface brightness limit & ICL fraction & r &  Surface brightness limit \\
& (\%) & (kpc) & (mag arcsec$^{-2}$) & (\%) & (kpc) & (mag arcsec$^{-2}$) \\
\hline
RXJ0856 & $9.96 \pm 2.72$ & 122 & $27.63 \pm 0.04$ & $14.41 \pm 0.93$ & 145 & $28.86 \pm 0.22$ \\
RXJ1136 & $3.77 \pm 1.98$ & 91 & $27.65 \pm 0.05$ & $6.06 \pm 2.84$ & 108 & $29.11 \pm 0.16$ \\
RXJ1410 & $4.73 \pm 2.39$ & 126 & $26.81 \pm 0.04$ & $10.68 \pm 1.81$ & 169 & $27.40 \pm 0.11$ \\
RXJ1007 & $7.24 \pm 3.48$ & 138 & $27.21 \pm 0.05$ & $12.39 \pm 0.50$ & 178 & $28.36 \pm 0.14$ \\
\hline
\end{tabular} 
\end{table}

\section{Discussion} \label{sect:discussion}

\begin{figure}
\includegraphics[width=1.\textwidth]{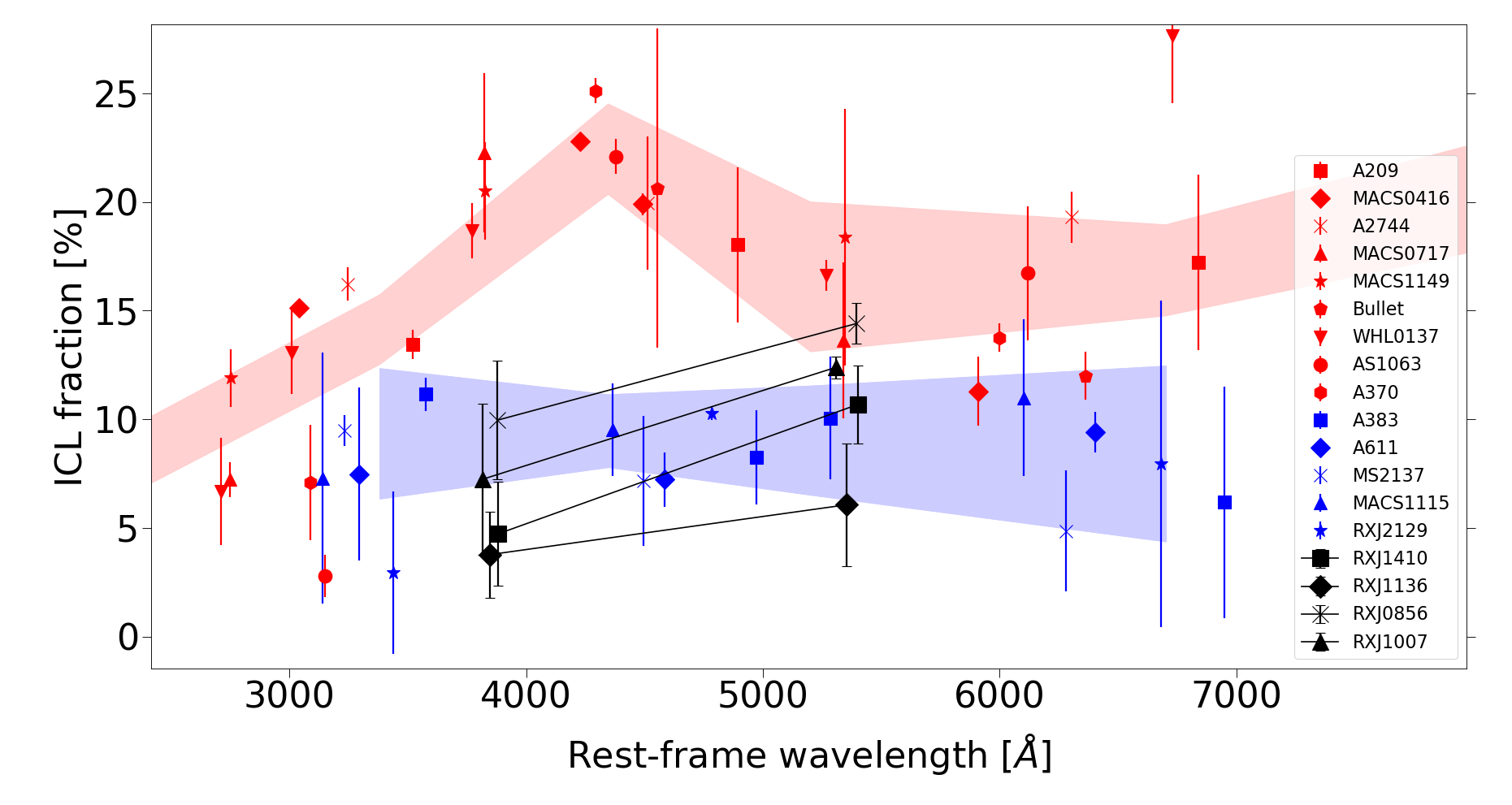}
\caption{ICL fractions measured for the three FGs added to a sample of galaxy clusters and groups analyzed by \cite{jimenez-teja2018,jimenez-teja2021} and \cite{deoliveira2022} at rest-frame wavelength. Merging systems are represented in red and relaxed systems are represented in blue. FGs systems are represented in black symbols with corresponding error bars. We also include the bona fide FG (RXJ1007, black triangle) analyzed by \cite{dupke2022}.} \label{fig:plot_iclf}
\end{figure}

In Fig. \ref{fig:plot_iclf}, we compare our ICL fractions with those measured for merging and relaxed clusters in previous works \citep{jimenez-teja2018,jimenez-teja2021,deoliveira2022} and for the FG RXJ1007 \citep{dupke2022}. All values are total ICL fractions, measured using HST/ACS data for clusters that span the redshift range $0.1 < z < 0.6$. Clusters classified as merging by different indicators are plotted with red symbols, while blue markers refer to relaxed systems. The shadowed red and blue regions indicate the error-weighted mean of the ICL fractions of merging and relaxed clusters, respectively. As shown in \cite{jimenez-teja2018}, relaxed clusters have constant ICL fractions (within the error bars) at different optical wavelengths, suggestive of a similar stellar composition for both the ICL and the cluster galaxies. This is the natural consequence of an ICL that is only fed by stars that are passively extracted from their progenitor galaxies while these orbit 
the center of the cluster (as is the case for relaxed systems). In contrast, the ICL fractions of active clusters show a characteristic peak between $\sim 3800-4800$ \AA~. This excess is provoked by a significant amount of young and/or low-metallicity stars that are violently thrown into the ICL in a relatively short period of time as a consequence of the galaxy mergers or near fly-bys, this mechanism being the main source of ICL production at these wavelengths.\\

\cite{dupke2022} pioneered the study of FGs via their ICL and showed that a bona fide FG, RXJ1007, had relatively low ICL fractions ($\sim 7-13$\%) in the wavelength interval $3800-5500$ \AA. Although these values were relatively consistent with those of relaxed clusters, they were slightly, but significantly, higher than usual for the F606W filter, i.e, at $\sim 5500$ \AA~ at rest-frame; see Fig. \ref{fig:plot_iclf}). This anomalous increasing trend suggested that this system was even more relaxed than regular relaxed clusters and contributed to estimate that the last merger event happened up to $\sim 6$ Gyr before. The behavior of the ICL fraction is consistent with that expected from a very relaxed system because:
\begin{enumerate}
\item the younger and/or lower-metallicity stars that are violently injected into the ICL during a merger had enough time to evolve so that their contribution in flux to the blue filter considered here decreases, 
\item later stellar types would be transferred from the galaxies to the ICL, since the only mechanism to produce ICL in an unperturbed and extremely passive system like a FG is the gradual tidal stripping of the cluster members while they approach the BCG. Given the observed radial gradient of metallicity and age found for nearby galaxies as one moves from the galaxy core outwards \citep[e.g.,][]{gonzalez-delgado2015}, the stars transferred from the galaxies to the ICL would be progressively redder,
\item Finally, all new star formation would happen preferentially within the galaxies, relatively increasing the ICL fraction in redder bands compared to bluer wavelengths.
\end{enumerate}

The ICL fraction difference between the F606W and F435W bands found for two of our three FG candidates, RXJ0856 ($\Delta f_{ICL}^{F606W-F435W}=4.45\pm2.87\%$) and RXJ1410 ($\Delta f_{ICL}^{F606W-F435W}=5.95\pm3.0\%$), are very similar to those of RXJ1007 ($\Delta f_{ICL}^{F606W-F435W}=5.15\pm3.5\%$). As expected, the ICL fractions do not show the distinctive peak found for merging clusters, have relatively low values, and show an increasing trend with wavelength. This result confirms the FG nature of these systems and suggests a common pattern for the ICL fractions of FGs at $z\sim 0.1$ in the wavelength range $\sim3800 - 5500$ \AA. However, while RXJ1007, RXJ0856, and RXJ1410 show a similar pattern in their ICL fractions, we do not see quantitatively the same behavior for those of RXJ1136 ($\Delta f_{ICL}^{F606W-F435W}=2.29\pm3.46\%$), whose distribution is constant within error bars (see Fig. \ref{fig:plot_iclf}).  Although this is indicative of a relaxed system, it is not enough to significantly mirror the behavior of the other FGs and may not even be a FG. Indeed, even though \cite{miller2012} classified RXJ1136 as a FG candidate based on photometric redshifts, our spectroscopic observations with GMOS allowed us to identify many new member galaxies. As a result, the magnitude gap measured between the brightest galaxy and the second-ranked galaxy in RXJ1136 within $0.5 R_{200}$ decreased to $\sim 0.73$ mag, which unambiguously breaks the main original criterion for classifying it as a FG.
In this case, both the ICL fractions and the spectroscopic redshift membership analysis agree that RXJ1136 is not a bona fide FG, although the former confirms that it is a relaxed system.\\

The cluster membership identification using new  spectroscopic redshifts (see Sect. \ref{sect:analysis}) for RXJ1410 shows that the magnitude gap between the first and second brightest galaxies within 0.5 $R_{200}$ is 2.26 mag, well above the threshold established in the classical definition of FGs. Contrarily, both the second and third ranked galaxies of RXJ0856 violate this criterion, with magnitude gaps of 1.83 and 1.88 mag, respectively. Although lower than the established threshold of 2 magnitudes, the photometric error associated with these measurements would make the magnitude gaps with the BGG compatible with a value of 2. Moreover, these two galaxies could be outside or right at the limit of 0.5 $R_{200}$, given that they are located at a projected distance of 621 and 594 kpc, respectively, and $R_{200}= 1.40\pm0.2$ Mpc for this cluster (see Table \ref{table:clusters}). This makes the classification of RXJ0856 as FG dubious if it is solely based on the traditional criteria of the magnitude gap, but clearly unambiguous from the ICL analysis.


\section{Conclusions} \label{sect:conclusions}

We performed the ICL analysis of three putative FGs, RXJ08564, RXJ1136, and RXJ1410, at z $\sim$ 0.1 using optical data in the F435W and F606W filters, obtained with HST/ACS, and spectroscopic data obtained with GMOS. We applied CICLE to generate the ICL maps and calculate the ICL fractions. We summarize the main results below:

\begin{enumerate}
\item The difference of the ICL fraction in the F606W and F435W bands for RXJ0856 and RXJ1410 is similar to that found for the first FG where this type of analysis was carried out, RXJ1007 \citep{dupke2022}, and is consistent with systems that are old, very relaxed perhaps having achieved the end of their merging trees.

\item 
No difference is seen between the ICL fraction in the F606W and F435W bands for RX1136 within the errors, which is similar to that found for normal relaxed clusters. Together with the recently observed violation of the magnitude gap criterion of FGs, with a value of 0.73 mag, this suggests that this system is not yet a FG.
\end{enumerate}

As the analysis of the ICL fraction provides a probe of the galactic system's dynamical state, the results presented in this work reinforce the ICL as a potential independent identifier of FGs beyond the traditional criteria.\\

\begin{acknowledgements}
N.O. acknowledges financial support from Conselho Nacional de Desenvolvimento Científico e Tecnológico (CNPq, Brazil) under the grant No. 141631/2020-1. Y.J-T. and J.M.V. acknowledge financial support from the State Agency for Research of the Spanish MCIU through Center of Excellence Severo Ochoa award to the Instituto de Astrofísica de Andalucía CEX2021-001131-S funded by MCIN/AEI/10.13039/501100011033, and from the grant PID2022-136598NB-C32 Estallidos and project ref. AST22-00001-Subp-15 funded from the EU-NextGenerationEU. RA acknowledges partial support support from CNPq grant 312565/2022-4 and NASA grants 80NSSC24K0484 and 80NSSC23K0291.

Based on observations obtained at the international Gemini Observatory, a program of NSF NOIRLab, which is managed by the Association of Universities for Research in Astronomy (AURA) under a cooperative agreement with the U.S. National Science Foundation on behalf of the Gemini Observatory partnership: the U.S. National Science Foundation (United States), National Research Council (Canada), Agencia Nacional de Investigaci\'{o}n y Desarrollo (Chile), Ministerio de Ciencia, Tecnolog\'{i}a e Innovaci\'{o}n (Argentina), Minist\'{e}rio da Ci\^{e}ncia, Tecnologia, Inova\c{c}\~{o}es e Comunica\c{c}\~{o}es (Brazil), and Korea Astronomy and Space Science Institute (Republic of Korea).

Funding for the Sloan Digital Sky Survey V has been provided by the Alfred P. Sloan Foundation, the Heising-Simons Foundation, the National Science Foundation and Participating Institutions. SDSS acknowledges support and resources from the Center for High-Performance Computing at the University of Utah. SDSS telescopes are located at Apache Point Observatory, funded by the Astrophysical Research Consortium and operated by New Mexico State University, and at Las Campanas Observatory, operated by the Carnegie Institution for Science. The SDSS web site is \url{www.sdss.org}.\\

SDSS is managed by the Astrophysical Research Consortium for the Participating Institutions of the SDSS Collaboration, including Caltech, The Carnegie Institution for Science, Chilean National Time Allocation Committee (CNTAC) ratified researchers, The Flatiron Institute, the Gotham Participation Group, Harvard University, Heidelberg University, The Johns Hopkins University, L’Ecole polytechnique f\'{e}d\'{e}rale de Lausanne (EPFL), Leibniz-Institut f\"{u}r Astrophysik Potsdam (AIP), Max-Planck-Institut f\"{u}r Astronomie (MPIA Heidelberg), Max-Planck-Institut f\"{u}r Extraterrestrische Physik (MPE), Nanjing University, National Astronomical Observatories of China (NAOC), New Mexico State University, The Ohio State University, Pennsylvania State University, Smithsonian Astrophysical Observatory, Space Telescope Science Institute (STScI), the Stellar Astrophysics Participation Group, Universidad Nacional Aut\'{o}noma de M\'{e}xico, University of Arizona, University of Colorado Boulder, University of Illinois at Urbana-Champaign, University of Toronto, University of Utah, University of Virginia, Yale University, and Yunnan University.
\end{acknowledgements}

\facilities{HST(ACS/WFC), Gemini:South(GMOS)}

\bibliographystyle{aasjournal}
\bibliography{main}{}

\end{document}